\documentclass[fleqn,twoside]{article}
\usepackage[headings]{espcrc2}

\readRCS
$Id: espcrc2.tex,v 1.2 2004/02/24 11:22:11 spepping Exp $
\usepackage{graphicx}
\usepackage[figuresright]{rotating}

\newcommand{\AmS}{{\protect\the\textfont2
  A\kern-.1667em\lower.5ex\hbox{M}\kern-.125emS}}

\usepackage{amsmath}
\usepackage{amssymb}
\usepackage{amsthm}
\usepackage{bbm}

\usepackage{maple2e}
 
\DefineParaStyle{Maple Output}
\DefineCharStyle{2D Math}
\DefineCharStyle{2D Output}
\DefineParaStyle{Bullet Item}
\DefineParaStyle{Heading 1}
\DefineParaStyle{Heading 2}
\DefineParaStyle{Warning}
\DefineCharStyle{2D Comment}
\DefineCharStyle{Help Normal}
\DefineCharStyle{Hyperlink}

\newcommand{\Gr}{Gr{\"o}bner }

\newcommand{\cJ}{\cal{J}}

\newcommand{\Z}{\mathbb{Z}}
\newcommand{\K}{\mathbb{K}}

\newcommand{\R}{\mathbb{R}}

\title{A Maple Package for Computing \Gr Bases for Linear Recurrence Relations}

\author{Vladimir P. Gerdt\address[JINR]{Laboratory of Information Technologies,
       Joint Institute for Nuclear Research,
       141980 Dubna, Russia}\thanks{gerdt@jinr.ru}
       and
       Daniel Robertz\address[LBFM]{Lehrstuhl B f{\"u}r Mathematik, RWTH Aachen,
       Templergraben 64, D-52062 Aachen,
       Germany}\thanks{daniel@momo.math.rwth-aachen.de}}

\runtitle{A Maple Package for Computing \Gr bases for Linear Recurrence Relations}
\runauthor{V. P. Gerdt and D. Robertz}

\begin{document}

\begin{abstract}
A Maple package for computing \Gr bases of linear difference ideals is described.
The underlying algorithm is based on Janet and Janet-like monomial divisions associated with
finite difference operators. The package can be used, for example, for automatic
generation of difference schemes for linear partial differential equations and for reduction of multiloop
Feynman integrals. These two possible applications are illustrated by simple examples
of the Laplace equation and a one-loop scalar integral of propagator type.
\end{abstract}

\maketitle

%
%
\section{INTRODUCTION}

As it is argued in~\cite{G04}, \Gr bases form the most universal
algorithmic tool for reduction of loop integrals determined by
linear recurrence relations derived from the integration by parts
method. These recurrence relations can be considered as generators
of an ideal in the ring of finite difference polynomials. Hence, to
determine the minimal set of basis (master) integrals and to
express other integrals in terms of basis integrals it suffices to
compute any \Gr basis, i.e., for any ranking (ordering) on
difference operators.

Another important application of difference \Gr bases is generation of
difference schemes for partial differential equations~\cite{MB01,GBM05}. A difference
scheme is constructed by elimination of partial derivatives from the system of difference
equations obtained by discretization of the differential equations and certain integral
relations. Therefore, the corresponding elimination ranking is to be used.

In this paper we present a Maple package for computing \Gr bases
of linear difference ideals. The package is a modified version
of our earlier package~\cite{Daniel} oriented towards commutative and
linear differential algebra and based on the involutive basis
algorithm~\cite{InvAlg}. The modified version is specialized to
linear difference ideals and uses both Janet and Janet-like
divisions~\cite{GB05} adopted to linear difference
polynomials~\cite{G-ACAT}.

%
%
\section{THE MAPLE PACKAGE LDA}

\subsection{Outline of the package}

The package LDA (abbreviation for {\bf{L}}inear {\bf{D}}ifference
{\bf{A}}lgebra) implements the involutive basis algorithm
\cite{InvAlg}\footnote{However, Janet trees for the fast search of
involutive divisors are not yet implemented since Maple does not
provide efficient data structures for trees.} for linear systems
of difference equations using Janet division. In addition, the package
implements a modification of the algorithm
oriented towards Janet-like division~\cite{GB05} and, thus, computes Janet-like
\Gr bases of linear difference ideals~\cite{G-ACAT}.

Table~\ref{table:LDA}
collects the most important
commands of LDA. Its main procedure {\tt JanetBasis} converts a
given set of difference polynomials into its Janet basis or
Janet-like \Gr basis form. More precisely, let $\R$ be the
difference ring \cite{G-ACAT} of polynomials in the variables
$\theta^\mu \circ y^k$, $\mu \in \Z_{\ge 0}^n$, $k = 1$, \ldots, $m$, with
coefficients in a difference field $\K$
containing $\mathbbm{Q}$ for which the field operations can be
carried out constructively in Maple. We denote by $\R_L$ the set of linear
polynomials in $\R$.
Given a generating set $F \subset \R_L$ for a
linear difference ideal $I$ in $\R$, {\tt JanetBasis}
computes the minimal Janet(-like Gr{\"o}bner) basis $J$ of $I$
w.r.t.\ a certain monomial order (ranking).
The input for {\tt JanetBasis} consists of the left hand
sides of a linear system of difference equations in the
dependent variables $y^1$, \ldots, $y^m$, e.g. functions of
$x_1$, \ldots, $x_n$. The difference
ring $\R$ is specified by the lists of independent variables
$x_1$, \ldots, $x_n$ and dependent
variables given to {\tt JanetBasis}. The output 
is a list containing the Janet(-like
Gr{\"o}bner) basis $J$ and the lists of independent and dependent
variables.

After $J$ is computed, the involutive/${\cJ}-$normal form of any element of
$\R_L$ modulo $J$ can be computed using {\tt InvReduce}.
Given $p \in \R_L$ representing a residue class $\overline{p}$ of
the difference residue class ring $\R / I$, {\tt InvReduce}
returns the unique representative $q \in \R_L$ of $\overline{p}$ which
is not involutively/${\cJ}-$reducible modulo $J$.
A $\K$-basis of the vector space $\R_L / (I \cap \R_L)$ is returned by
{\tt ResidueClassBasis} as a list if it is finite or is enumerated by
a formal power series \cite{JanetsApproach}
in case it is infinite. For examples of how to apply
these two commands, cf.\ Section~\ref{section:Feynman}.

Given an affine (i.e. inhomogeneous) linear system of difference
equations, a call of {\tt CompCond} after the application of
{\tt JanetBasis} returns a generating set of
compatibility conditions for the affine part of the system,
i.e. necessary conditions for the right hand sides of the
inhomogeneous system for solvability.

Moreover, combinatorial devices to compute the
Hilbert series and polynomial and function etc. \cite{Daniel}
are included in LDA.

For the application of LDA to the reduction of Feynman
integrals, a couple of special commands were implemented
to impose further relations on the master integrals:
By means of {\tt AddRelation} an infinite sequence
of master integrals parametrized by indeterminates which
are not contained in the list of independent variables
is set to zero. Subsequent calls of {\tt InvReduce}
and {\tt ResidueClassBasis} take these additional
relations into account (cf.\ Section~\ref{section:Feynman}).

Finally, LDA provides several tools for dealing with
difference operators. Difference operators represented by
polynomials can be applied to (lists of) expressions containing
$y^1$, \ldots, $y^m$
as functions of the independent variables.
Conversely, the difference operators can be extracted
from systems of difference equations.
Leading terms of difference equations
can be selected.

\begin{table*}[htb]
\caption{Main commands of LDA}
\label{table:LDA}
\renewcommand{\tabcolsep}{2pc} 
\renewcommand{\arraystretch}{1.2} 
\begin{tabular}{@{}ll}
\hline
{\tt JanetBasis}                  & Compute Janet(-like Gr{\"o}bner) basis\\
{\tt InvReduce}                   & Involutive / ${\cJ}-$reduction modulo Janet(-like Gr{\"o}bner) basis\\
{\tt CompCond}                    & Return compatibility conditions for inhomogeneous system\\
{\tt HilbertSeries} etc.          & Combinatorial devices\\
{\tt Pol2Shift}, {\tt Shift2Pol}  & Conversion between shift operators and equations\\
\hline
\multicolumn{2}{@{}l}{Some interpretations of commands for the reduction of Feynman integrals:} \\
\hline
{\tt ResidueClassBasis}           & Enumeration of the master integrals\\
{\tt AddRelation}                 & Definition of additional relations for master integrals\\
{\tt ResidueClassRelations}       & Return the relations defined for master integrals\\
\hline
\end{tabular}
\end{table*}

\subsection{Implementation details}

We consider difference rings containing
shift operators which act in one direction only.
If a linear system of difference equations is given
containing functions shifted in both directions,
then the system needs to be shifted by the
maximal negative shift in order to obtain a
difference system with shifts in one direction only.
However, LDA allows to change the shift direction
globally.

Unnecessary computations of involutive reductions
to zero are avoided using the four involutive criteria
described in \cite{Criteria}, \cite{UnnecessaryReduction}.
Fine-tuning is possible by selecting the criteria
individually.

The implemented monomial orders/rankings are the
(block) degree-reverse-lexicographical
and the lexicographical one. In the case of more
than one dependent variable, priority of comparison
can be either given to the difference operators
(``term over position'') or to the dependent variables
(``position over term''/elimination ranking).

The
ranking is controlled via options given to
each command separately. The other options described above
can be set for the entire LDA session using
the command {\tt LDAOptions} which also allows
to select Janet or Janet-like division.

%
%
\section{GENERATION OF FINITE DIFFERENCE SCHEMES FOR PDES}

%

We consider the Laplace equation $u_{xx} + u_{yy} = 0$
and rewrite it as the conservation law
\[
\oint_{\Gamma} -u_y dx + u_x dy = 0.
\]
Adding the integral relations
\begin{eqnarray*}
\int_{x_j}^{x_{j+2}} u_x dx & = & u(x_{j+2}, y) - u(x_j, y),\\
\int_{y_k}^{y_{k+2}} u_y dx & = & u(x, y_{k+2}) - u(x, y_k)
\end{eqnarray*}
and using the midpoint integration method we obtain
the following discrete system:
\begin{equation} \label{eq:LaplaceDiscrete}
\left\{
   \begin{array}{l}
    -(\theta_x - \theta_x \theta_y^2) \! \circ \! u_y  +
(\theta_x^2 \theta_y - \theta_y) \! \circ \! u_x = 0, \\
    2 \triangle h\, \theta_x \! \circ \! u_x   - (\theta_x^2 - 1) \! \circ \! u = 0,   \\
    2 \triangle h\, \theta_y \! \circ \! u_y  - (\theta_y^2 - 1) \! \circ \! u = 0,
   \end{array}
 \right.
\end{equation}
where $\theta_x$ and $\theta_y$ are the right-shift operators w.r.t. $x$ and $y$, e.g.,
$\theta_x\circ u_y(x,y)=u_y(x+1,y)$.

We show how to use LDA to find a finite difference scheme
for the Laplace equation:

\medskip

\begin{maplegroup}
\begin{mapleinput}
\mapleinline{active}{1d}{with(LDA):}{%
}
\end{mapleinput}
\end{maplegroup}

\medskip

\noindent
We enter the independent and the dependent variables
for the problem ($ux > uy > u$):

\medskip

\begin{maplegroup}
\begin{mapleinput}
\mapleinline{active}{1d}{ivar:=[x,y]: dvar:=[ux,uy,u]:}{%
}
\end{mapleinput}
\end{maplegroup}

\medskip

\noindent
Next, we translate (\ref{eq:LaplaceDiscrete}) into the input format
of {\tt JanetBasis}. Note that one can in general use
{\tt AppShiftOp} to apply a difference operator given
as a polynomial similar to the ones in (\ref{eq:LaplaceDiscrete})
to a difference polynomial.

\medskip

\begin{maplegroup}
\begin{mapleinput}
\mapleinline{active}{1d}{L:=[2*h*ux(x+1,y)-u(x+2,y)+u(x,y),
2*h*uy(x,y+1)-u(x,y+2)+u(x,y),
2*h*(ux(x+2,y+1)-ux(x,y+1))+
2*h*(uy(x+1,y+2)-uy(x+1,y))]:}{%
}
\end{mapleinput}
\end{maplegroup}

\medskip

\noindent
Then we compute the minimal Janet basis of the linear difference ideal
generated by $L$ w.r.t. a ranking which compares the
dependent variables prior to the corresponding difference monomials
(``position over term'' order; this ranking is chosen when
using the option $2$ as below).
The least element of this Janet
basis is by construction a difference polynomial which does not
contain any monomial in $ux$ and $uy$ because $ux > uy > u$.

\medskip

\begin{maplegroup}
\begin{mapleinput}
\mapleinline{active}{1d}{JanetBasis(L,ivar,dvar,2)[1][1];}{%
}
\end{mapleinput}

\mapleresult
\begin{maplelatex}
\mapleinline{inert}{2d}{u(x+4,y+2)-4*u(x+2,y+2)+u(x,y+2)+u(x+2,y+4)+u(x+2,y);}{%
\maplemultiline{
 \mathrm{u}(x + 4, \,y + 2) - 4\,\mathrm{u}(x + 2, \,y + 2) + \mathrm{u}(x,
 \,y + 2) \\
 + \mathrm{u}(x + 2, \,y + 4) + \mathrm{u}(x + 2, \,y) }
}
\end{maplelatex}
\end{maplegroup}

\medskip

The computation takes less than one second of time on a
Pentium III (1 GHz).

Dividing this difference polynomial by $4h^2$ we obtain the
following finite difference scheme:
\[
D_j^2(u_{jk})+D_k^2(u_{jk})=0,
\]
where
\[
D_j^2(u_{jk})=\frac{u_{j+2 \, k} - 2 u_{j \, k} + u_{j-2 \, k}}{4
h^2}
\]
and
\[
D_k^2(u_{jk})= \frac{u_{j \, k+2} - 2 u_{j \, k} + u_{j \, k-2}}{4
h^2}
\]
are discrete approximations of the second order partial derivatives
occurring in Laplace's equation.



%
%
\section{REDUCTION OF FEYNMAN INTEGRALS} \label{section:Feynman}

%
%

In order to demonstrate how to use LDA for the reduction of
Feynman integrals, we consider a simple one-loop propagator type
scalar integral with one massive and another massless particle:
\[
 f(k,n):=I_{k, n} = \frac{1}{i\pi^{d/2}}
 \int  \frac{d^d s}
  {P_{s-q,m}^{k} P_{s,0}^{n}}.
\]
(Here $k$, $n$ are the exponents of the propagators.)

The basis integrals for this example and the corresponding
reduction formulae were found and studied by several authors (see,
e.g.,~\cite{Tarasov,Smirnov}). Here we apply the \Gr basis
method, as implemented in LDA, directly to the
recurrence relations which have the form:
\begin{equation} \label{eq:FeynmanRecurrence}
\left\{ \begin{array}{l}
\, \! \! [d-2k-n - 2m^2 k \mathbf{1}^{+} -\\
\; \! \! n\mathbf{2}^{+}(\mathbf{1}^{-}-q^2+m^2)] \, f(k+1,n+1)=0,\\
\, \! \! [n-k - k \mathbf{1}^{+}(q^2+m^2-\mathbf{2}^{-}) -\\
\; \! \! n\mathbf{2}^{+}(\mathbf{1}^{-}-q^2+m^2)] \, f(k+1,n+1)=0,
\end{array} \right.
\end{equation}
where
\[
\mathbf{1}^{\pm}f(k,n)=f(k\pm 1,n),\
\mathbf{2}^{\pm}f(k,n)=f(k,n\pm 1).
\]
In addition, it is known that
\begin{equation} \label{eq:AdditionalRelation}
f(k+i,n+j)=0 \quad \forall \, i \leq 0 \quad \forall \, j
\end{equation}
which we will take into account later.

\medskip

\begin{maplegroup}
\begin{mapleinput}
\mapleinline{active}{1d}{ivar:=[k,n]: dvar:=[f]:}{%
}
\end{mapleinput}
\end{maplegroup}

\medskip

\noindent
We enter the recurrence relations (\ref{eq:FeynmanRecurrence}):

\medskip

\begin{maplegroup}
\begin{mapleinput}
\mapleinline{active}{1d}{L:=[(d-2*k-n)*f(k+1,n+1)-2*m^2*k*
f(k+2,n+1)-n*f(k,n+2)-n*(m^2-q^2)*
f(k+1,n+2),  (n-k)*f(k+1,n+1)-
k*(q^2+m^2)*f(k+2,n+1)+k*f(k+2,n)-
n*f(k,n+2)-n*(m^2-q^2)*f(k+1,n+2)]:}{%
}
\end{mapleinput}
\end{maplegroup}

\begin{maplegroup}
\begin{mapleinput}
\mapleinline{active}{1d}{JanetBasis(L,ivar,dvar):}{%
}
\end{mapleinput}
\end{maplegroup}

\medskip

\noindent
Again, the computation time is less than one second.
Now, the master integrals are given by:

\medskip

\begin{maplegroup}
\begin{mapleinput}
\mapleinline{active}{1d}{ResidueClassBasis(ivar,dvar);}{%
}
\end{mapleinput}

\mapleresult
\begin{maplelatex}
\mapleinline{inert}{2d}{[f(k,n), f(k,n+1), f(k+1,n), f(k,n+2), f(k+1,n+1), f(k+2,n)];}{%
\maplemultiline{[\mathrm{f}(k, \,n), \,\mathrm{f}(k, \,n + 1), \,
\mathrm{f}(k + 1, \,n),\\
\, \mathrm{f}(k, \,n + 2), \,\mathrm{f}(k + 1, \,n + 1), \,
\mathrm{f}(k + 2, \,n)]}
}
\end{maplelatex}
\end{maplegroup}

\medskip

\noindent
(\ref{eq:AdditionalRelation}) implies additional relations on
the master integrals. (Here, $j$ is recognized as not being
contained in {\tt ivar} and thus serves as a parameter to
define the additional relations.)

\medskip

\begin{maplegroup}
\begin{mapleinput}
\mapleinline{active}{1d}{AddRelation(f(k,n+j)=0,ivar,dvar):}{%
}
\end{mapleinput}
\end{maplegroup}

\medskip

\noindent
The list of master integrals now becomes:

\medskip

\begin{maplegroup}
\begin{mapleinput}
\mapleinline{active}{1d}{ResidueClassBasis(ivar,dvar);}{%
}
\end{mapleinput}

\mapleresult
\begin{maplelatex}
\mapleinline{inert}{2d}{[f(k+1, n), f(k+1,n+1), f(k+2,n)];}{%
\[
[\mathrm{f}(k + 1, \,n), \,\mathrm{f}(k + 1, \,n + 1), \,\mathrm{f}(k + 2, \,n)]
\]
}
\end{maplelatex}
\end{maplegroup}

\medskip

\noindent
Next, we recompute the Janet basis for $m=0$:

\medskip

\begin{maplegroup}
\begin{mapleinput}
\mapleinline{active}{1d}{m:=0: J:=JanetBasis(L,ivar,dvar):}{%
}
\end{mapleinput}
\end{maplegroup}

\medskip

\noindent
For the special case where $m = 0$, we impose the relation
$f(k+i, n) = 0$ for all $i$:

\medskip

\begin{maplegroup}
\begin{mapleinput}
\mapleinline{active}{1d}{AddRelation(f(k+i,n)=0,ivar,dvar):}{%
}
\end{mapleinput}
\end{maplegroup}

\medskip

\noindent
Now, we are left with one master integral:

\medskip

\begin{maplegroup}
\begin{mapleinput}
\mapleinline{active}{1d}{ResidueClassBasis(ivar,dvar);}{%
}
\end{mapleinput}

\mapleresult
\begin{maplelatex}
\mapleinline{inert}{2d}{[f(k+1,n+1)];}{%
\[
[\mathrm{f}(k + 1, \,n + 1)]
\]
}
\end{maplelatex}
\end{maplegroup}

\medskip

\noindent
We reduce $f(k+2,n+3)$ modulo $J$ taking also
the additionally imposed relations on the master
integrals into account. (Here, the option ``F''
lets {\tt InvReduce} return the result in
factorized form.)

\medskip

\begin{maplegroup}
\begin{mapleinput}
\mapleinline{active}{1d}{InvReduce(f(k+2,n+3),J,"F");}{%
}
\end{mapleinput}

\mapleresult
\begin{maplelatex}
\mapleinline{inert}{2d}{-(2*n+4-d+2*k)*(2*n+2-d+2*k)*(2*k+n-d)*(n+3-d+k)*(n+2-d+k)*f(k+1,n+1
)/((n+1)*(2*n-d+4)*n*q^6*k*(d-2*k-2));}{%
\maplemultiline{
 - ((2\,n + 4 - d + 2\,k)\,(2\,n + 2 - d + 2\,k) \\
(2\,k + n - d)\,(n + 3 - d + k) \\
(n + 2 - d + k)\,\mathrm{f}(k + 1, \,n + 1))/( (n + 1)\\
(2\,n - d + 4)\,n\,q^{6}\,k \,(d - 2\,k - 2)) }
}
\end{maplelatex}
\end{maplegroup}

\medskip

\noindent
Using {\tt ResidueClassRelations} one can display
the relations imposed on the master integrals:

\medskip

\begin{maplegroup}
\begin{mapleinput}
\mapleinline{active}{1d}{ResidueClassRelations(ivar,dvar,[i,j]);}{%
}
\end{mapleinput}

\mapleresult
\begin{maplelatex}
\mapleinline{inert}{2d}{[f(k,n+j), f(k+i,n)];}{%
\[
[\mathrm{f}(k, \, n + j), \,\mathrm{f}(k + i, \,n)]
\]
}
\end{maplelatex}
\end{maplegroup}

\medskip

\noindent
The difference operators occurring in the last result can
be extracted as polynomials in $\delta_k$, $\delta_n$:

\medskip

\begin{maplegroup}
\begin{mapleinput}
\mapleinline{active}{1d}{Shift2Pol(\%,ivar,dvar,
[delta[k],delta[n]]);}{%
}
\end{mapleinput}

\mapleresult
\begin{maplelatex}
\mapleinline{inert}{2d}{[delta[n]^j, delta[k]^i];}{%
\[
[{\delta _{n}}^{j}, \,{\delta _{k}}^{i}]
\]
}
\end{maplelatex}
\end{maplegroup}



%
%
\section{CONCLUSION}

We presented the Maple package LDA implementing the Janet/Janet-like
division algorithm for computing \Gr bases of linear difference ideals. Generation
of difference schemes for linear PDEs and reduction of loop Feynman integrals are
important applications of the package.

These two kinds of applications were illustrated by rather simple
examples. The first difference system (discrete Laplace equation
and integral relations) contains two independent variables $(x,y)$
and three dependent variables $(u,u_x,u_y)$. The second system
(recurrence relations for one-loop Feynman integral) also contains
two independent variables/indices $(k,n)$, but the only dependent
variable $f$. The second system, however, is computationally
slightly harder than the first one because of explicit dependence
of the recurrence relations on the indices and three parameters
$(d, m^2, q^2)$ involved in the dependence on indices.

Dependence on index variables and parameters is an attribute of recurrence
relations for Feynman integrals. Similar dependence may occur in the generation of
difference
schemes for PDEs with variable coefficients containing parameters. Theoretically
established exponential and superexponential (depends on the ideal and ordering)
complexity of constructing polynomial \Gr bases implies
that construction of difference \Gr bases is at least exponentially hard in the number of
independent variables (indices). Besides, in the presence of parameters
the volume of computation grows very rapidly as the number of parameters increases.

Thus, for successful application of the \Gr basis technique to multiloop
Feynman integrals with masses and to multidimensional PDEs with multiparametric
variable coefficients we are going not only to improve our Maple code
but also to implement the algorithms for computing Janet and/or Janet-like difference bases in
C++ as a special module of the open source software available on the Web page {\tt http://invo.jinr.ru}.

\section{ACKNOWLEDGEMENTS}

The authors are grateful to Vladimir Smirnov
for useful
discussion, comments and suggestions on the example of
Section~\ref{section:Feynman}. The contribution of the first
author (V.P.G.) was partially supported by the grants 04-01-00784
and 05-02-17645 from the Russian Foundation for Basic Research and
the grant 2339.2003.2 from the Russian Ministry of Science and
Education.

\end{document}